\documentclass[letterpaper,aps,superscriptaddress,floatfix,twocolumn]{revtex4}

\usepackage{graphicx}
\usepackage{hyperref}
\usepackage{amsmath}
\usepackage{url}
\usepackage{subfigure}
\usepackage{color}

\usepackage{times}
\begin{document}

\title{Accelerator Based Fusion Reactor}

\author{Keh-Fei Liu}{\Large {\Large }}
\affiliation{Department of Physics and Astronomy, University of
Kentucky, Lexington, Kentucky 40506, USA}

\author{Alexander W. Chao}
\affiliation{SLAC National Accelerator Laboratory, Stanford University, Stanford, CA 94309}

\begin{abstract}
A feasibility study of fusion reactors based on accelerators is carried out. We consider a novel scheme where 
a beam from the accelerator hits the target plasma on the resonance of the fusion reaction and establish
characteristic criteria for a workable reactor. We consider  the reactions $ d + t \rightarrow n + \alpha,
d + {}^3H_e \rightarrow p + \alpha$, and $p + {}^{11}B \rightarrow 3 \alpha$ in this study. The critical temperature 
of the plasma is determined from overcoming the stopping power of the beam with the fusion energy gain. The needed plasma 
lifetime is determined from the width of the resonance, the beam
velocity and the plasma density. We estimate the critical beam flux by balancing the energy 
of fusion production against the plasma thermo-energy and the loss due to stopping power for the case of
an inert plasma. The product of critical flux and plasma lifetime is independent of plasma density and has a weak dependence
on temperature. Even though the critical temperatures for these reactions are lower than those for the thermonuclear reactors, the critical 
flux  is in the range of $10^{22} - 10^{24}/\rm{cm^2/s}$ for the plasma density $\rho_t = 10^{15}/{\rm cm^3}$ in the case
of an inert plasma. Several approaches to control the growth of the two-stream instability are discussed. We have also
considered several scenarios for practical implementation which will require further studies.
Finally, we consider the case where the injected beam at the resonance energy maintains the plasma temperature
and prolongs its lifetime to reach a steady state. The equations for power balance and particle number
conservation are given for this case. 
\end{abstract}


\maketitle

     Harnessing energy from controlled fusion reaction has been a challenge for more 
than six decades. In the intervening years, great progress has been made toward attaining sufficient 
confinement time and density at the required temperature to sustain a net yield of energy from the
fusion reaction. This has culminated in the ITER program which is designed to produce 500 MW 
power sustained up to 1000 s with an energy gain of a factor of $\sim 10$~\cite{ITER}.

Despite much progress made in thermonuclear reactor and inertial confinement, the time scale
of commercial production is still far off. There are alternative designs to Tokamak being explored, such as 
stellarator~\cite{STELLARATOR}, field  reversed configurations (FRC)~\cite{FRC,Guo15} and dense 
plasma focus (DPF)~\cite{DPF,Lerner12}. 
Why do we want to consider yet another scheme involving accelerators?
The thermonuclear reactor such as at ITER is at a temperature (12.5 keV) which is much lower than that
of the  peak resonance energy of the $d+t$ reaction with a center of mass energy of 64 keV. Thus, it is the exponential
tail of the Maxwell-Boltzmann distribution that is important in the integrated reaction rate $\langle \sigma v\rangle
\sim 10^{-22}\,{\rm m^3/s}$. Whereas,  direct $d$ on $t$ on the resonance yields a $\langle \sigma v\rangle
= 1.6 \times 10^{-21} m^3/s$ which is an order of magnitude larger.  In fact, most of the light ion fusion reactions 
have resonances at center-of-mass energy of 64 -- 300 keV with widths of 200 -- 400 keV. It would be reasonable 
to ask if one can take advantage of this feature and explore the possibility of a fusion reactor with the fusion nuclei 
colliding at the energy where the fusion cross section peaks in order to maximize the reaction rate. This will require
a beam at a particular energy. Consequently, an accelerator is a logical tool in this regard.

   However, the reaction rate (i.e. reactivity) is not the only concern for a reactor to work. All the possible energy losses
need to be taken into account. In the straightforward approach to the accelerator based fusion reactor (ABFR), where the
beam from the acceleator is used as the fuel, there 
can be insurmountable difficulties. For example, simply bombarding the target at room temperature with a beam
will not work. The ratio of fusion energy gain as denoted by the $Q$-value vs stopping power 
\mbox{$R_e = \sigma Q/(|dE/dx|/\rho_t) = 0.013$} is much less than unity for 100 - 200 keV proton 
on Helium~\cite{stoppower}. In other words, the stopping power due to the bound electrons in atoms, which includes ionization and beam bremsstrahlung, overwhelms the fusion energy production. There has been a design to consider
colliding beams in storage rings~\cite{rug00} with beam density at $\sim 10^{15}/{\rm cm^3}$. However, the transverse
momentum impulse due to Coulomb repulsion at 1mm from the center for a beam size of $1\, {\rm cm}^2$, is an order
of magnitude larger than the beam longitudinal momentum~\cite{Sands:1970ye}. Thus, the beams will splash sideways 
instead of going through each other to initiate fusion reaction. Neutral beam shields long range Coulomb interaction, but the cross section for ionization, such as 
\mbox{$\sigma (H_2 + H_e \rightarrow H_2 + H_e^+ + e)|_{100\,{\rm keV}} = 5 \times 10^{-17} {\rm cm}^2$} is 7 orders of magnitude larger than that of the fusion cross section, so that the energy loss due to ionization is much larger than the fusion energy gain. One can also consider the laser wakefield setup~\cite{TD79} where the electrons are temporarily pushed to
the rim of the bubble in the plasma by the laser and separated from the ions in the blow-out region. In this case, one can guide the beam into the blow-out region when the bubble is formed to avoid interaction with the electrons. However, the characteristic time
scale of the bubble lifetime of pico-sec is too short for the non-relativistic beam with velocity of 1 - 3\% of the
speed of light to go through the bubble. 

    Notwithstanding the above examples which illustrate various difficulties of ABFR with or without electrons around, one notices that it is possible to have electrons around and yet be innocuous. This brings us to our proposal of making the plasma the target. Ion beams
on plasma have been considered before, but not for ABFR. Neutron beam injection~\cite{kms72} has been utilized to heat up the plasma in Tokamak~\cite{gri12} and FRC~\cite{Guo15} reactors. Heavy ion beams have been considered a promising driver option for fast ignition in inertial confinement facilities~\cite{mas75,lin95}. In the present work, we shall consider the external beam from the accelerator as the fuel itself for the fusion energy  production in ABFR for the first time. The primary reason for considering
this arrangement is to take advantage of a specific, perhaps unique, feature of the plasma in that the stopping powers of the beam due to the electrons and ions in the plasma decrease with temperature as $T^{-3/2}$~\cite{Chu72,Hamada78}. Therefore, 
by raising the temperature of the plasma, sooner or later the energy loss due to stopping power will yield to fusion gain. 
We shall
first consider the simplified case of an inert plasma, by which we mean the plasma in a volume $V$ has a constant density $\rho_t$, constant temperature $T$, and a lifetime of $\tau_{pla}$ as given parameters and there is no dynamical response to the incoming beam. In this case, the net energy gain is
\begin{equation}    \label{energy-balance}
\Delta E_{\rm net} = E_{\rm fus} - E_{\rm sp} - E_{\rm pla} > 0,
\end{equation}
where $E_{\rm fus}$ is the fusion energy production
\begin{equation}
E_{\rm fus} = \rho_b  v \rho_t V \sigma Q \tau_{\rm pla} \epsilon_{\rm out},
\end{equation}
with $\rho_b/v$ being the beam density/velocity.  $Q$ is the energy gain. $\sigma$ is the fusion reaction cross section. 
$\tau_{pla}/\rho_t$ is the plasma lifetime/density. 
$\epsilon_{out}$ is the output energy conversion efficiency to electricity. For charged particle production, the direct conversion is possible which gives $\epsilon_{out} \sim 0.9$. For neutron production, $\epsilon_{out} \sim 0.3-0.4$. We take  0.3
for the present work. The energy loss $E_{\rm sp}$ 
due to stopping power is
\begin{equation}
E_{\rm sp} = \rho_b  v \rho_t V (Z\,|dE/ dx|/\rho_e) \tau_{\rm pla}/\epsilon_b,
\end{equation}
where $|dE/dx|/\rho_e $ is the stopping power per unit target electron density and we have used $\rho_e = Z \rho_t $ for the
neutral plasma where $Z$ is the charge of the ions.
 $\epsilon_b$ is the energy efficiency of producing the beam. High efficiency klystron for proton linac sources at proton energy of
 115 keV has reached an efficiency of 65\%~\cite{Beu13}. The overall energy efficiency for the beam will be lower. We shall take  
 $\epsilon_b = 0.5$ as a working number for the present work. The thermo-energy loss of the plasma during $\tau_{pla}$ is
\begin{equation}
E_{\rm pla} =  \rho_t V n_{\rm eff}(3/2\, T)/\epsilon_{pla},
\end{equation}
where $n_{\rm eff}$ is the effective number of charged particles per ion in the plasma which is 6 for ${}^{11}B$,
3 for ${}^3H_e$ and 2 for $t$, assuming equipartition. $\epsilon_{pla}$ is the efficiency for producing
the plasma. We take it to be 0.5 in this work. In general, one can consider the scenario where the electron and ion temperatures are different. In the present work, we shall consider them to be the same.

We note that Eq.~(\ref{energy-balance}) can be written as
\begin{eqnarray}    \label{energy-balance-2}
\Delta E_{\rm net} &=& \rho_t V \tau_{\rm pla}\Big\{\phi \big[\,\overline{\sigma Q} \epsilon_{\rm out} 
- \frac{Z\,|dE/dx|}{\ \epsilon_b\,\rho_e} \big]   \nonumber \\
&-& n_{\rm eff}(3/2 T)/(\tau_{\rm pla}\epsilon_p)\Big\} > 0,
\end{eqnarray}
where $\phi = \rho_b v$ is the beam flux density and $\overline{\sigma Q}$ is the average of $\sigma Q$. We shall take it
to be the average between $E_R + \Gamma/2$ and  $E_R - \Gamma/2$ with $\overline{\sigma Q} \sim 3/4\, \sigma_{\rm max} Q$,
where $E_R/\Gamma$ is the resonance energy/width of the fusion reaction.
As we see from Eq.~(\ref{energy-balance-2}),
besides having to overcome $E_{\rm pla}$,  the expression inside the square bracket should be larger than zero so that the fusion energy 
production could offset the loss of stopping power. This leads to

\vspace*{0.2cm}
$\bullet$ {\it Criterion 1:}
\vspace*{-0.25cm}
\begin{equation}   \label{R_sp}
R_{\rm sp} = \frac{\overline{\sigma Q} \epsilon_{\rm out}}{Z\,|dE/dx|/(\rho_e\epsilon_b)} 
=  \frac{\overline{\sigma Q}\rho_e \, v\, \epsilon_{\rm out}}{Z|dE/dt|/\epsilon_b} \ge 1.
\end{equation}
%
The second equality in Eq.~(\ref{R_sp}) is just the ratio of fusion power production vs. the power loss to
the stopping power for each beam particle with the efficiencies $\epsilon_{\rm out}$ and $\epsilon_b$ taken into account.
The stopping power of plasma in the quantum regime with $T \ge 1$ keV  for non-relativistic ions goes down with the plasma temperature as $T^{-3/2}$~\cite{Chu72} and is proportional to $v$, i.e. $|dE/dx|/\rho_e \propto  v\,T^{-3/2}$ with a logarithmic correction~\cite{Hamada78}. An exact calculation with quantum correction to the order of
$g^2 \ln g^2$ is given~\cite{Brown:2005ji} for the plasma coupling $g = e^2 \kappa_D/4\pi T$ where $\kappa_D$ 
is the Debye wave number. A comparison of the stopping power for proton with speed $v_p = 0.0365 c$ in the plasma with 
$\rho_e = 5 \times 10^{25}/{\rm cm^3}$ and $T = 1$ keV to that at $\rho_e =  10^{24}/{\rm cm^3}$ and $T = 0.2$ keV shows
that the $T^{-3/2}$ scaling is good to $\sim$ 20\% and there is an approximate $v$ scaling. 
While a more precise calculation will be given later, we shall adopt the $v\,T^{-3/2}$ scaling for the present study with
the proviso that it is good to a factor of 2 for the range of $T$ and beam velocity in this work.  We take the proportionality constant from proton at $v_p/c = 0.0365$ which will produce the $p + {}^{11} B \rightarrow 3\, \alpha$ reaction at maximum
cross section and obtain 
\vspace{-0.45cm}
\begin{equation}  \label{stopping-T}
|dE/dx|/\rho_e  = a\, v\, T^{-3/2} 
\end{equation}
where $a = 7.27 \times 10^{-29}{\rm \,keV\,cm\,s\,(keV)}^{3/2}$ is from Ref.~\cite{Brown:2005ji}.

    Following criterion 1 in Eq.~(\ref{R_sp}), we determine the critical temperature $T_c$ at $R_{sp} =1$ which is
\mbox{$T_c = (a Z v/(\epsilon_b \epsilon_{\rm out}\,\overline{\sigma Q}))^{2/3}$} and tabulate it in
 \mbox{Table~\ref{tab:T_c}} for three reactions. 

 Next, we determine the lifetime and the length of the plasma in order to maximize the fusion reaction
with the beam energy entering the plasma at $E_R  + \Gamma/2$ and exiting at $E_R - \Gamma/2$.
The energy loss is due to the stopping power, therefore we have the effective length of the plasma to be
\begin{equation}    \label{lmax}
l_{\rm eff} = \frac{\Gamma}{|dE/dx| (\rho_e, T)}.
\end{equation}
Here, the stopping power $|dE/dx| (\rho_e, T)$ depends on the electron density and the temperature of the plasma. 
The parameters of the three fusion reactions, such as the resonance energy ($E_R$) in the lab frame,
the width of the resonance ($\Gamma$), the projectile velocity $v/c$ at $E_R$,
the Q value, the maximum fusion cross section ($\sigma_{\rm max}$) and the incoming beam
energy ($E_b$) at $E_R + \Gamma/2$ are given in Table~\ref{tab:T_c}.  
We shall consider a scenario for  the low density at $\rho_t = 10^{15}/{\rm cm}^3$ which is relevant to the characteristic
density of Tokamak~\cite{ITER}, stellarator~\cite{STELLARATOR}, and FRC~\cite{FRC,Guo15} and a high density one
at $\rho_t = 10^{21}/{\rm cm}^3$ which is achievable in DPF~\cite{Lerner12,BBB95}.

Taking the plasma lifetime  $\tau_{\rm pla}$ 
to be the beam traverse time, we obtain it from Eqs.~(\ref{stopping-T}) and (\ref{lmax})
\begin{eqnarray}   \label{tau_pla}
\tau_{\rm pla} \equiv \tau_{\rm tra} = l_{\rm eff}/v  = \frac{\Gamma\, T^{3/2}}{a\, v^2 \, Z\, \rho_t}.
\end{eqnarray}
We note that $\tau_{pla}$ increases with $T$ as $T^{3/2}$ and is inversely proportional to $\rho_t$. 
We list the $\tau_{pla}$ in Table~\ref{tab:T_c} which is in the range of $10^{-2}/10^{-7} s$ for the low-/high-density scenarios.  

We should point out that $\tau_{pla}$ is not a criterion, it is the desired
plasma lifetime that would maximize the fusion reaction with the traversing beam. On the other hand, the maximum useful lifetime
of the plasma is when the fuel of the incoming beam is used up. We estimate this by dividing the mean-free-path of the 
beam particle by its speed, i.e. $\tau_{\rm max} \sim 2/(\rho_t \sigma_{\rm max} v)$. To compare with $\tau_{\rm pla}$, we look at 
the ratio 
$\frac{\tau_{\rm pla}}{\tau_{\rm max}} = \frac{\Gamma \sigma_{\rm max} T^{3/2}}{2 a Z v}$,
%
which, at $T_c$, equals $2 \Gamma/(3 \epsilon_b \epsilon_{out} Q)$ which is 0.033, 0.035, and 0.051 for the
$d\,t, d\,{}^3\!H_e$ and $p\,{}^{11}\!B$ reactions. This means that the beam particles lose energy faster than they burn out
through the fusion reaction. Since the ratio is less than unity, there is room for $\tau_{\rm pla}$ to be longer than those 
at $T_c$. It can be achieved by increasing $T$ 
until it reaches a maximum 
$T_{\rm max}$ where the ratio becomes unity. In this case, $T_{max} = (2a Z v/(\Gamma \sigma_{max}))^{2/3}$. 
We tabulate this $T_{\rm max}$ in Table~\ref{tab:T_c} also.

\begin{widetext} 
  \begin{center}
 \begin{table}[tb]
  \centering
  \renewcommand{\arraystretch}{1.4}
  \caption{Critical plasma temperature for criterion in Eq.~(\ref{R_sp}),  $l_{e\!f\!f}$, and 
  $\tau_{pla}$ for two scenarios of the plasma density. 
  Other relevant parameters, i.e. the resonance energy $E_R$, the width $\Gamma$, the beam speed $v/c$, the $Q$ value, the maximum
  fusion cross-section $\sigma_{\rm max}$, and the beam energy $E_b$ are also tabulated for reference.}
  \begin{tabular} {|c|cccccccc|cc|cc|}
  \hline
  &&&&&&&&&
   \multicolumn{2}{|c|}{$\rho_t=10^{15}/{\rm cm^3}$}  & \multicolumn{2}{c|}{$\rho_t=10^{21}/{\rm cm^3}$} \\
   \cline{10-13}
  Reaction & $T_c$(keV) & $T_{\rm max}$(keV) & $E_R $(keV) & $\Gamma$(keV) & $v/c$ & Q (MeV) &  
  $\sigma_{\rm{\rm max}}$(b) &
  $ E_{\rm{b}}$(keV) & $l_{\rm eff}$(cm) & $\tau_{\rm pla}$ (s)  &  $l_{\rm eff}$(cm) & $\tau_{\rm pla}$ (s)  \\
    \hline
     d + t                 & 1.8   &  17 & 160 & 210 & 1.07\% & 17.6 & 5.1 & 265 & 1.3 $\times 10^{7}$ & $4.0 \times 10^{-2}$ & 13 & 4.0 $\times 10^{-8}$ \\
    \hline
    d +${}^3H_e$     & 7.0  & 66 & 438 & 430 & 2.16\% & 18.4 & 0.81   & 653  & $8.5 \times 10^7$   & $1.3 \times 10^{-1}$ &  85   &  $ 1.3 \times 10^{-7}$ \\
    \hline
    p  + ${}^{11}$B     &  31  & 223 & 625 & 300 & 3.65\% & 8.7  &  0.80    &  775   &  $1.3 \times 10^8$   & $1.2 \times 10^{-1}$    &  $1.3  \times     10^{2}$  & $1.2\times 10^{-7}$  \\
    \hline
      \end{tabular}   \label{tab:T_c}
 \end{table}
 \end{center}
 \end{widetext}
We see from Table \ref{tab:T_c} that the plasma lifetime and $l_{\rm eff}$ for $\rho_t=10^{21}/{\rm cm^3}$ are commensurate with those achievable in DPF for the high-density scenario.  
For the low-density scenario, we find that while $\tau_{\rm pla}$ is not a problem, the linear dimension of
$l_{\rm eff}$ at $\sim 10^5 {\rm m}$  is too long for the size of a practical linear reactor. However, there is no need to be
limited to a linear reactor with this length. One can consider curvilinear trajectories of the beam. We will discuss this later.
%

For the next step, we consider energy balance for the case of an inert plasma. We see from Eq.~(\ref{energy-balance-2}) that, to gain 
net energy,  not only should the fusion energy gain offset the loss in the stopping power, it should also overcome the energy to 
produce the plasma with a certain lifetime. According to Eqs.~(\ref{energy-balance-2}) and (\ref{tau_pla}), the critical minimal beam flux $\phi_c$ is determined by

%
\begin{eqnarray}  \label{phi_c}
\phi_c = \!\frac{3/2\, n_{\rm eff}\, Z\, T}{\epsilon_{\rm out}\, \epsilon_{\rm pla}\, \overline{\sigma Q}\, g(T)\, \tau_{\rm pla}}
   = \!  \frac{3/2\, n_{\rm eff}\, Z\, a\, v^2\, \rho_t}{\epsilon_{\rm out}\, \epsilon_{\rm pla}\, \overline{\sigma Q}\, g(T) \,T^{1/2}},
\end{eqnarray}
where $g(T) = 1 - |dE/dx|/(\epsilon_{\rm out}\, \epsilon_{\rm pla}\, \overline{\sigma Q})$. 
%

 \begin{figure}[htb]
\centering
\subfigure
{\includegraphics[width=1.0\hsize]{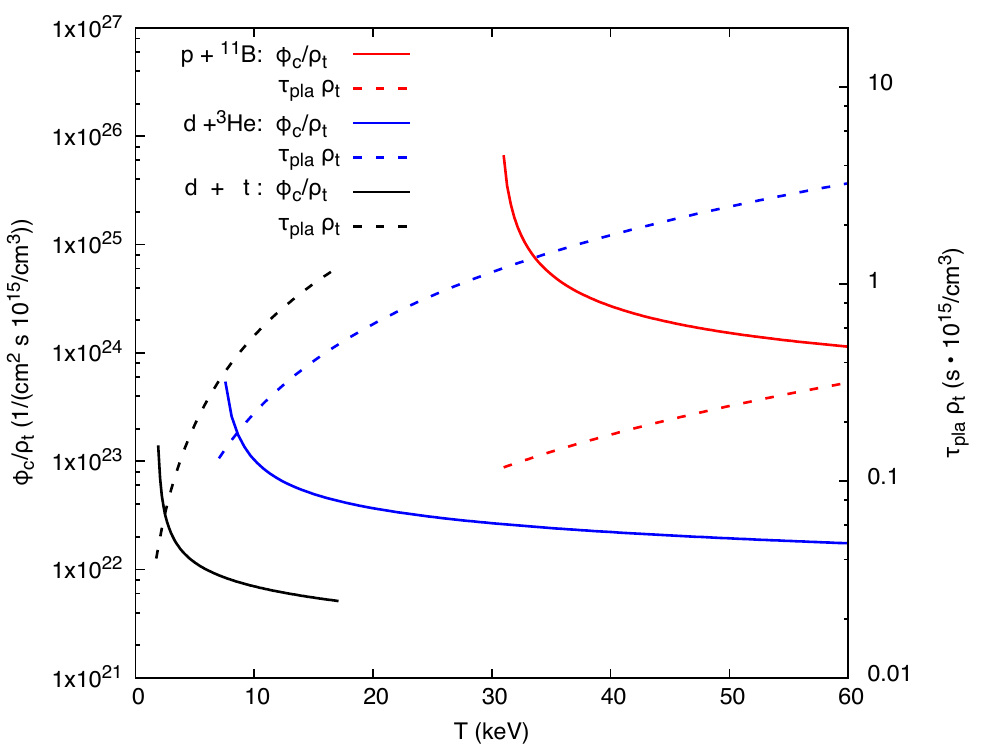}}
\subfigure
{\includegraphics[width=0.85\hsize]{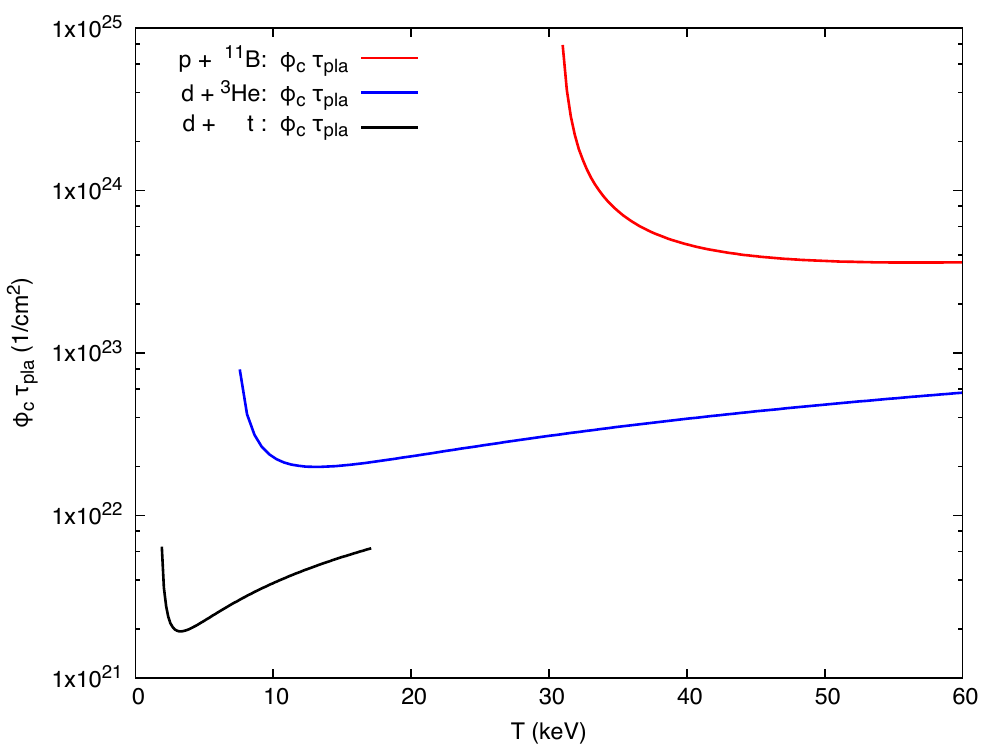}}
\caption{The critical flux divided by the plasma density  $\phi_c/\rho_t$ in units of  $ 1/({\rm cm^2\,s\,10^{15}/cm^3})$ and the associated plasma lifetime times its density $\tau_{\rm pla}\, \rho_t$ in units of ${\rm s\cdot10^{15}/cm^3}$ are plotted as a function of temperature for the three reactions in the upper panel. They terminate at $T_{\rm max} = 17.1$ keV for the $d + t$ reaction. $\phi_c\, \tau_{\rm phy}$ is plotted as a function 
of $T$ for the three reactions in the lower panel.} 
\label{fig:phi_tau}
\end{figure}

    Since $\phi_c$ is proportional to $\rho_t$ and $\tau_{\rm pla}$ is inversely proportional to $\rho_t$,
 we plot $\phi_c /\rho_t$ in units of $1/({\rm cm^2\,s\,\cdot 10^{15}/{cm}^3})$ and $\tau_{\rm pla}\, \rho_t$ in units of ${\rm s \cdot 10^{15}/cm^3}$ in 
 Fig.~\ref{fig:phi_tau} as a function of temperature.  Also plotted in the lower panel of Fig.~\ref{fig:phi_tau} is  
 %
 $ \phi_c\, \tau_{\rm pla} =  \frac{3/2\, n_{\rm eff} T}{\epsilon_{\rm out}\epsilon_{\rm pla} \overline{\sigma Q} g(T)}$,
 which is independent of $\rho_t$ and has minima at $T$ only a little higher than $T_c$ for the three reactions. 
 
 We note that the critical temperatures $T_c$ are lower than those needed in themonuclear reactors. For example, 
$T_c = 1.8$ keV for the $d + t$  case is lower than the target ITER temperature of 12.5 keV by a factor of 6.6. Similarly, 
\mbox{$T_c = 7.0/31$ keV} for the $d\,H_e/p\,B$ case is much smaller than the corresponding temperature of $\sim 60/300$ keV for 
ignition in the thermonuclear reactor~\cite{Nevins98}. This is because, unlike the case of the thermonuclear reactor,
ABFR aims to maximize $\langle \sigma v \rangle$ by colliding the beam on resonance.  
As a consequence, when one considers the triple product $\rho_t\, T_c\, \tau_{\rm pla} = 7.2 \times 10^{19}\, {\rm keV\, s/m^3}$ for the $d\, t$ reaction with 
$\rho_t = 10^{15}/{\rm cm^3}$, it is smaller than the Lawson criterion which is 
$\sim 3 \times 10^{21}\, {\rm keV\, s/m^3}$~\cite{Lawson57}. However, this is not quite a germane comparison. Even though the 
constraint on the triple product is lessened, the challenge is shifted to the demand of high beam flux for ABFR at 
$\rho_t = 10^{15}/{\rm cm^3}$.

 We see from the upper panel of Fig.~\ref{fig:phi_tau} that at the minima of  $\phi_c\, \tau_{\rm pla}$ (lower panel), the critical flux is between $10^{22}/({\rm cm^2s})$ and  $10^{24}/({\rm cm^2s})$  for the low  density scenario  (i.e.\ $10^{15}/{\rm cm^3})$. This is several orders of magnitude higher than the typical flux from linear accelerators with radio frequency quadrupole, such as $1 {\rm A}/{\rm cm^2} \sim10^{19}/({\rm cm^2s})$  in the first accelerating component at Spallation Neutron Source in ORNL~\cite{Cou15}. A simple solution is to lower $\rho_t$ by a few orders of magnitude at the expense of a proportionally increased $\tau_{\rm pla}$ (Eq.~(\ref{tau_pla})). Other possibilities include compression of the beam at injection which can reach $\sim 100 {\rm A/cm^2}$
 at this energy range~\cite{Zel14} , $H^{-}$ charge-exchange accumulator, cooling, and ultimately injection with multiple beams. The pros and cons of the various approaches or the combination of approaches should be studied during the design stage.
  
   So far, we have considered the requirements for the beam with an inert plasma. When the dynamical response of the plasma to the beam is taken into account, one of the concerns is that when the energetic beam is injected to the plasma, it can cause two-stream 
 instability~\cite{Mikh74} where the imaginary part of the frequency emerges so that the amplitude of the plasma oscillation and the
 electrostatic potential can grow exponentially.  It was pointed out that for ion beams on plasma, the phase velocity
of the waves in question is typically much smaller than that of the electrons. In this case, the latter can be treated as a neutralizing background fluid~\cite{Fri66}. This applies to our case since the phase velocity of the wave is of the order of the beam velocity 
at $1 \sim 3\%$ of $c$; whereas the thermal velocity of the electrons at 3 keV is $13\%$ of $c$. For the case that the plasma frequency 
of the beam ions is much smaller than that of the plasma ions, i.e. $\omega_{pb} \ll \omega_{pt}$, the dispersion relation of the linearized fluid equations can have complex roots for the frequency with the maximum growth rate $\gamma_{t} =  \frac{\sqrt{3}}{2^{4/3}} (\frac{\omega_{pb}}{\omega_{pt}})^{2/3}\, \omega_{pt}$~\cite{Mikh74}. Take, for example, the $d + t$ reaction with parameters from Table~\ref{tab:T_c} for  $\rho_t = 10^{15}/cc$, for which $\omega_{pb} \ll \omega_{pt}$ is satisfied, the maximum growth rate is $\gamma_{t} = 6.0 \times 10^{9} {\rm rad/s}$. For the duration of the plasma time from Table~\ref{tab:T_c}, the exponential growth factor is $\gamma_t \tau_{pla} =  2.4 \times 10^{8}$.
This is so large that it would render the ABFR considered so far with a monotonic continuous beam unfeasible. 
One way to ameliorate the growth rate is through the compression of the charged beam~\cite{SD06} and is confirmed in an NDCX experiment~\cite{Roy05}. Another possibility is to consider a pulsed beam to prevent the buildup of the exponential growth due to 
the continuous beam. Perhaps the most effective way to evade the two-stream instability is to have a spread in the beam velocity. 
Since the fusion reaction resonance is broad with a width of $200 - 400$ keV (see Table~\ref{tab:T_c}), a beam with a spread much less than the resonance width will not affect the reactivity much. The stability limits of longitudinal Langmuir waves in ion beam-plasma interaction has been studied~\cite{Fri66,Hay15}. For the case that $V = v_b/v_{th,i} \gg 1$ where $v_b/v_{th,i}$ is the 
beam/thermal plasma ion velocity, the stability limit for $\alpha = v_{th,i}/v_{th,b}$ which is the ratio of plasma ion velocity
to the thermal velocity of the beam $v_{th,b}$ is $\alpha_{max} = \sqrt{2(n+1)T/n} V - 1$ where $n= \rho_b/\rho_i$ is the dimensionless beam density and $T = T_i/T_e$ which is unity in the present study with equal ion and electron temperatures. For our
low-density scenario with critical $\phi_c$ with $V = 7.8, 7.9, 6.4$ for the $d+t, d + {}^3H_e, p+{}^{11}$B reactions, this requires   minimum beam temperatures of $0.90, 16, 230$ eV respectively, which are much smaller than those of the beam energies. When the ion beam streams through plasma, there is also a Weibel instability~\cite{wei59} which causes the magnetic field to grow exponentially with the rate $\gamma_w = (v_b/c)\, \omega_{pb}$. Since $v_b/c \sim 1\%$ and $\omega_{pb} < \omega_{pt}$, we have $\gamma_w \ll \gamma_t$, i.e. it is much smaller than the growth rate of the two-stream instability. Thus, the above discussed approaches to avoid the two-stream instability should also apply to avoid the Weibel instability.

Specific reactor designs, particularly detailed engineering designs, are beyond the scope of the present work. They will require experimental tests and numerical simulations. In view of the fact that accelerator technology is mature, its parameters for the beam, such as the beam energy, flux and bunching can be better controlled and, furthermore, they are decoupled from those of the plasma, one can consider the parameters of the plasma and the accelerator separately. This affords the opportunity to consider a range of different setups. We shall consider a few scenarios where we discuss the potential
caveats and challenges for future original design references. From Table~\ref{tab:T_c}, we see that for high plasma density 
($\rho_t = 10^{21}/cc$), the effective length of the plasma $l_{\rm eff}$ can be as short as 10 cm to 1 m, which can be a good size for the plasma. However, the required $\phi_c$ in this case will be $\sim 10^{27}/{\rm cm^2\, s}$ for d+t. This is 8 orders of magnitude larger than the typical beam flux in use, e.g. at SNS~\cite{Cou15}. It is not feasible with today's technology for a single beam. For the low density case ($\rho_t = 10^{15}/cc$),  the required $l_{\rm eff}$ from Table~\ref{tab:T_c} is $\sim 10^7 {\rm cm}$. This is too large to be  practical for a plasma device with this linear dimension. However, the charged beam can move in a circle in a constant perpendicular magnetic field so that its trajectory can be confined in a limited space region to be compatible with the physical size of the plasma~\cite{Crawford}. One scenario is to embed the plasma in a straight section of  a storage ring for the beam. 
A magnetic field perpendicular to the plane of the ring outside the plasma device is applied so that the beam bends in a circular path of the ring outside the plasma section. If the emittance growth of the beam due to Coulomb scattering of the plasma~\cite{Mon85} (N.B. even though the plasma is neutral on the average, there is still Coulomb scattering due to the fact that the charge is not neutral locally) is under control such that the beam can be recollected with high efficiency after it passes through the plasma section and continued on in the ring with perpendicular magnetic field so that it can be sent back and passed through the plasma multiple times. 
The strength of the magnetic field is determined from the Larmor radius which is the radius of the ring. For a ring with 10 m in radius, the magnetic field $B = \frac{m_b v}{|q| r} = 65\, {\rm}G$ is needed for $d$ beam with $v/c = 1.07\%$
(see Table~\ref{tab:T_c}) for the $d+t$ reaction. We shall consider the $d+t$ reaction for the following scenarios. For the plasma, one needs to distinguish two situations. One is the case where the plasma lifetime is relatively long. From Table~\ref{tab:T_c}, the required $\tau_{pla}$ is $\sim 40$ ms and $T_c = 1.8$ keV. This long lifetime needs plasma confinement. If these plasma parameters can be met with the magnetic mirror device (Note the recent experiment with a magnetic mirror device has reached an electron temperature of 900 eV and a lifetime longer than 8 ms~\cite{Bag15}), one can consider a storage ring of radius 10 m with the magnetic mirror in a straight section of the ring where the beam is to go through the axial direction of the mirror and interact with the plasma. In the middle section of the mirror where the magnetic field is parallel to the longitudinal axis, it has little effect on the beam. The challenge lies at the ends of the mirror where the magnetic field may have non-vanishing and non-uniform vertical components and will disperse the beam. The feasibility of this design will depend on how large the beam size is affected and if the emittance growth can be tolerated so that the beam can be collected back into the ring after it passes through the magnetic mirror device. The feasibility of this scenario can be explored with numerical simulation and experiments. Another case involving the storage ring is when the actual lifetime of the plasma is much shorter than $\tau_{tra}$ which is $\tau_{pla}$ (see Eq.~(\ref{tau_pla})), the required plasma lifetime for the beam to traverse the distance of  $l_{\rm eff}$. It has been demonstrated that a picosecond laser can produce a plasma from a  gas of hydrogen clusters at the density of $10^{15}/cc$ with the electron temperature over 5 keV for more than 200 ns~\cite{Dit97}. The energy absorption efficiency can be as high as 90\%~\cite{Dit97}. A similar condition can be reached with dense plasma focus (DPF)~\cite{Lerner12}. For such a short lifetime, there is no need to consider plasma confinement. During the lifespan of 200 ns, the beam will travel a distance of 64 cm. Consider a ring with a circumference of 12.8 m which can be divided into 20 sections. Each section can be injected with tritium gas clusters to be irradiated by the laser in time for the beam to traverse. After the beam circles once around the ring under the
appropriate magnetic field, it can be directed to a separate storage ring through a figure 8 configuration for example. The beam will be stored in this second ring and can be redirected back to the first ring when the next plasma is produced. While in the second ring without plasma, one can accelerate the beam with RF to regain the energy lost through the stopping power in the plasma ring so that the fusion reaction can be kept on resonance. This will depend on being able to recollect the beam after circling around the plasma ring with high efficiency. The net power generation will be degraded in this storage ring reactor compared to the ideal situation where a continuous beam passes through the plasma with the length $l_{\rm eff}$ for the duration of the required $\tau_{pla}$. The ratio of the power $P_{\rm sto}$ of this storage ring reactor to the ideal $P_{\rm ideal}$ is $P_{\rm sto}/P_{\rm ideal} = r_{\!f}\, \tau_{pla}  C/l_{\rm eff}$, where $r_{\!f}$ is the repetition frequency of the laser and $C$ is the circumference of the plasma ring. Taking $r_{\!f}$ to be 10 Hz and $ C = 12.8\, {\rm m}$, we find $P_{\rm sto}/P_{\rm ideal} = 4.1 \times 10^{-5}$ from the numbers in Table~\ref{tab:T_c}. For a beam of the size of  0.1 cm in radius, it will be shown later that $P_{\rm ideal} = 29\, {\rm MW}$. In this case, one has $P_{\rm sto} = 1.2 \,{\rm kW}$, adequate for the energy supply of a household. To design a higher power plant, one can increase the beam size (or have multiple beams in the storage ring pipe), the ring circumference, and the repetition frequency with multiple lasers. Neutral beam injection was developed in the late seventies and is now one of the main heating methods for most of the fusion experiments, such as the Tokamak~\cite{ITER}, the Stellerator~\cite{STELLARATOR}, and FRC~\cite{FRC} devices. One can conduct experiments by sending a neutral deuterium beam into the magnetically confined tritium plasma right above the resonance with large flux $\phi_c$ to see if the idea of ABFR is feasible for this arrangement.

    There have been proposals to consider non-Maxwellian plasmas where the energy of the alpha particles from the fusion reaction is transferred to the light ions to form a monoenergetic beam to increase reactivity and lower the ignition temperature~\cite{Hay15}. However, it is not clear how this
is to be realized practically and how the two-stream instability is to be controlled. 

     To the extent that the two-stream and other instabilities are under control,
the temperature maintained by the injected beam and charged fusion products and the lifetime extended to have
the fuel burned out, then the stringent requirement on the beam flux as prescribed in Eq.~(\ref{phi_c}) can be reduced. This is because
the stopping power is not totally lost, part of it will heat up the plasma or maintain its temperature in a steady state. 
If the non-Maxwellian plasma idea works, it can be adopted to rekindle the reactivity.
 In this case, $g(T)$ in Eq.~(\ref{phi_c}) is closer to unity which serves to decrease $\phi_c$. 
 When and if such a steady state is achieved, both the power balance and nucleon number conservation are required. This brings up  
 a more realistic criterion when the total system including the dynamical response of the plasma is taken into account. 
 
 \vspace*{0.2cm}
$\bullet$ {\it Criterion 2:}
\vspace*{0.0cm}
\begin{eqnarray}   \label{steady_state}
P_{\rm fus} + P_{\rm reheat} &=&   P_{\rm sp} + P_{\rm b+pla} + P_{\rm rad} + P_{\rm cond} + P_{\rm leak},  \nonumber \\
\frac{dN_b}{d t} + \frac{d N_{\rm pla}}{dt} & =&  \frac{dN_{\rm fus+leak}}{dt} + \frac{dN_{\rm lowE}}{dt},
\end{eqnarray}
where $P_{\rm fus}$ is the power generated by fusion and  $P_{\rm reheat}$ is the power of plasma reheating due to the
transferring of kinetic energy from the stopping power loss of the beam.
$P_{\rm b+pla}$ is the power of the beam and plasma.
 $P_{\rm rad}/P_{\rm cond}$ is the power loss due to radiation/conduction. 
$P_{\rm leak}$ accounts for the particles leaking from the plasma including neutrons and energetic charged particles which are not confined in the plasma. Similarly, $d N_b/d t$ and  $d N_{\rm pla}/dt$ are the
rate of supply of the fuel from the beam and the plasma; whereas, $ d N_{\rm fu+leak}/dt$ is the rate 
of producing fusion products as well as particle leakage and $d N_{\rm lowE}/dt$ is the rate for increasing those beam particles which lose enough energy so that they are far below the resonance region to be eligible for fusion reaction.

 As for power generation, it depends on many factors. To give an order of magnitude estimate, we take the plasma temperature to be
 close to the minimum of $\phi \tau_{pla}$ and $\phi = \phi_c$ at this temperature, and assume that
 the plasma repetition frequency is commensurate with $\tau_{\rm pla}$. Therefore, 
 $P_{\rm net} = \frac{\Delta E_{\rm net}}{\tau_{\rm pla}} = \rho_t\,A\, v\, n_{\rm eff}(3/2 T)/\epsilon_p$,
 where $A$ is the beam cross-section. $P_{\rm net}$ depends linearly on $\rho_t$, $A$, and the plasma repetition fequency in this case.  Given $\rho_t = 10^{15}/{\rm cm^3}$ and the radius of the beam size to be 0.1 cm, the power generated by the three reactions are  29 MW ($d\,t$, $T = 3$ keV), 0.35 GW ($d\,H_e$, $T= 12$ keV) and 5.1 GW ($p\,B$, $T = 50$ keV). It can be scaled up by
 increasing the beam size or $\rho_t$. On the other hand, to scale it down to kW range, one can consider decreasing the 
 repetition frequency of the plasma and beam supplies or the plasma density by several orders of magnitude. 
 It has been concluded that the Bremsstrahlung loss in inertial electrostatic confinement (IEC) and thermonuclear systems
 is prohibitively large for $p + ^{11}$B reactor. We find that the electron Bremsstrahlung loss rate $P_{\rm Brem}$
 are 0.95 MW ($d\,t$), 95 MW ($d\,H_e$), and 4.9 GW ($p\,B$). They are smaller than their respective $P_{\rm net}$.
 The  reason that the Bremsstrahlung problem for  $p + ^{11}$B reactor is evaded here is due to the fact that the temperature
 is lower and the fusion reactivity $\langle \sigma v\rangle$ larger than those in the thermonuclear reactor.

     In summary, we have considered the feasibility of fusion reactors based on a novel approach of using the beam from accelerators as the fuel to be injected into a plasma at the resonance energies. We set up a first criterion on the 
 critical temperature by considering the stopping power of the beam in the plasma and the fusion energy production. They turn  
 out to be several times smaller than those needed for the thermonuclear reactor for the three reactions we considered.
 Considering an inert plasma, we estimated the minimum plasma lifetime and beam flux from the resonance width and energy balance.  For the more realistic case of a dynamical plasma, we considered several approaches including pulsed beam and a beam of modest temperature to avoid the two-stream instability and presented criteria due to energy and nucleon number conservations for a steady state of such a reactor. 
 In this new approach, the parameters of the accelerator and the plasma are decoupled, this adds additional dimensions to the traditional thermonuclear reactors and has the potential of enriching the possibility of innovative designs of fusion reactors incorporating accelretators.   Exploring this advantage of the ABFR approach, we have considered practical implementation and discussed the caveats and challenges in several scenarios. Each will require further studies and experimentation. 
 
We thank M. Cavagnero, S. Cousineau, C. Crawford, \mbox{T. Draper}, Wei Lu, and J.S. Zhang for discussions and encouragement.

\end{document}